\documentclass[preprint,3p,twocolumn,10pt]{elsarticle}
\usepackage[english]{babel}

\usepackage{booktabs}
\usepackage{multirow}
\usepackage{mathtools}
\usepackage{xcolor}
\usepackage{gensymb}
\usepackage{subfig}

\usepackage{amssymb}

\journal{Physical review C}

\begin{document}

\begin{frontmatter}

\title{High precision measurement of the half-life of the 391.6\,keV metastable level in $^{239}\textrm{Pu}$}

\author[uno,due]{A.~Barresi\corref{cor}}

\cortext[cor]{Corresponding authors: \newline massimiliano.nastasi@unimib.it, \newline a.barresi@campus.unimib.it}

\author[uno,due]{D.~Chiesa}
\author[uno,due]{M.~Nastasi\corref{cor}}
\author[uno,due,tre]{E.~Previtali}
\author[due]{M.~Sisti}
\address[uno]{Department of Physics, University of Milano-Bicocca, 20126 Milan, Italy}
\address[due]{INFN, Milano-Bicocca, 20126 Milan, Italy}
\address[tre]{INFN, Laboratori Nazionali del Gran Sasso, Assergi (L'Aquila) I-67100 - Italy}

\begin{abstract}
Materials selection for rare event physics requires high performance detectors and customized analyses. In this context a novel $\beta$-$\gamma$ detection system, comprised of a liquid scintillator in coincidence with a HPGe, was developed with the main purpose of studying ultra trace contamination of uranium, thorium and potassium in liquid samples. In the search for $^{238}\textrm{U}$ contaminations through neutron activation analysis, since the activation product $^{239}\textrm{Np}$ decays to a relatively long-lived isomeric state of $^{239}\textrm{Pu}$, it is possible to perform a time selection of these events obtaining a strong background suppression.
Investigating the time distribution of the coincidences between the $\beta^-$ decay of $^{239}\textrm{Np}$ and the delayed events following the de-excitation of the $^{239}\textrm{Pu}$ isomeric level at 391.6\,keV, a precision measurement of the half-life of this level was conducted. The half-life of the 391.6 keV $^{239}\textrm{Pu}$ level resulted $190.2 \pm 0.2$ ns, thus increasing the precision by about a factor of 20 over previous measurements.
\end{abstract}


\end{frontmatter}



\section{Introduction}
\label{S:1}
\begin{figure}[ht!]
	\begin{center}
		\includegraphics[width=0.48\textwidth]{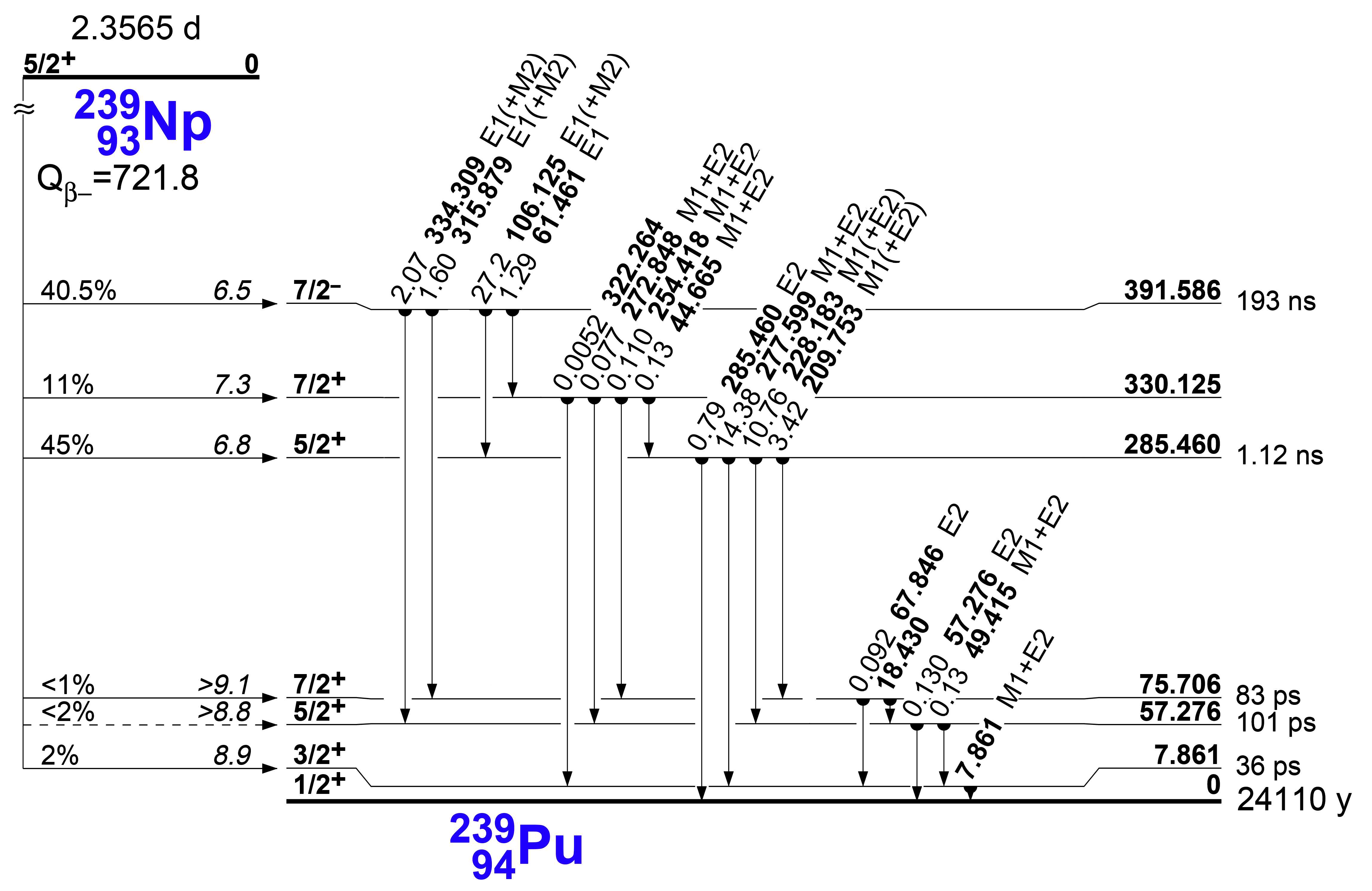}
		\caption{Simplified nuclear level scheme of $^{239}\textrm{Pu}$\cite{TOI}. The main transitions involved in the measurement of the half-life of the metastable level are shown.}
		\label{fig:decayscheme}
	\end{center}
\end{figure}

Since the 1950's the $^{239}\textrm{Np}$ nuclear structure has been studied extensively \cite{Baranov,Borner,Ewan,Patel,Mackenzie} leading to a well-known level scheme.
As shown in Figure \ref{fig:decayscheme}, $^{239}\textrm{Np}$ $\beta^-$ decay populates with high probability a metastable level of $^{239}\textrm{Pu}$ at 391.6 keV above the ground state level. Consequently this level decays to the lower energy states with delayed $\gamma$ emission or internal conversion transitions (IC) \cite{Ewan,Mackenzie}. In 1955 Engelkemeir and Magnusson, by exploiting a coincidence circuit between anthracene and sodium iodide scintillation counters,  performed the measurement of the half-life of $^{239}\textrm{Pu\,(391.6\,keV)}$ level, achieving a result of $193 \pm 4$ ns \cite{Engelkemeir}. Almost twenty years ago, S.B. Patel et al. confirmed the same result for $\textrm{T}_{1/2}$ of the 391.6~keV level: $192\pm6\,\textrm{ns}$ \cite{Patel}, by studying the electromagnetic properties of the excited states of $^{239}\textrm{Pu}$. Both results are statistically consistent, but with the best uncertainty of only 4~ns (2\%). 

An accurate knowledge of the half-life of $^{239}\textrm{Pu\,(391.6\,keV)}$ level has a specific implication in ultra sensitive measurements of $^{238}\textrm{U}$. In the context of materials selection for rare events physics experiments, neutron activation analysis (NAA) is a good tool to determine ultra trace of contamination of $^{238}\textrm{U}$. This kind of analysis is usually performed by measuring an irradiated sample with high purity germanium detectors (HPGe) in low background configuration. Interfering nuclides within activated sample represent a limit in this approach, since they create a background which could overlap the signal of interest. Exploiting the delayed emission from the 391.6 keV level, a time-based analysis allows to identify events emitted from $^{239}\textrm{Pu}$, removing random coincidence generated by the background or by interfering activated isotopes. Experimental advantages from these considerations could be achieved by developing a custom detector that allows to perform a time-based analysis of the events. 
This paper describes how the half-life of the 391.6~keV level was determined with an uncertainty of 0.2~ns (0.1\%). This result was achieved through the delayed coincidence measurement between the $\beta^{-}$ and the electrons produced by IC transitions or photon interactions by photoelectric or Compton effects in the LS (IC/$\gamma$ electrons) generated by the deexcitation cascade of the metastable level.
The measurement was performed by a system made of a liquid scintillator (LS) and HPGe detectors, named GeSparK. This detector was primarily developed to determine the radioactive contamination of activated liquid samples. Thanks to its design it allows to measure the half-life of metastable levels whose half-life is long enough compared to the time response of the LS detector.

In the following sections the experimental setup, the measurement leading principles, the data analysis and measurement results, and the evaluation of the systematic errors are discussed.

\section {Detector description}

In the context of low background radioactivity measurements, a new detector, GeSparK, was developed in the Radioactivity Laboratory of the Department of Physics of the University of Milano-Bicocca\cite{GeSparK}. It is a composite system consisting of a liquid scintillator sealed in a Teflon container coupled to a photomultiplier tube (PMT), and a HPGe detector working in time coincidence, thus allowing the acquisition of decay events characterized by well-defined time correlations. A dedicated acquisition system allows to digitize the signals from both the LS and HPGe detectors in a specific time window. In accordance with its structure and thanks to the excellent time resolution of the LS detector (few ns), the GeSparK system can identify $\alpha$-$\gamma$ and $\beta$-$\gamma$ coincidence events, rejecting all the events which do not respond to the requested temporal features. This capability drastically reduces the environmental and cosmogenic backgrounds, thus improving the analytical sensitivity. 

\section {Measurement principle}

Thanks to its particular setup, GeSparK detector allows to perform a very accurate measurement of the half-life of the $^{239}\textrm{Pu}$(391.6 keV) metastable state. 
As shown in Figure \ref{fig:decayscheme}, $^{239}\textrm{Np}$ decays with a 40.5\% branching ratio on that level with a $\beta^-$ transition which is followed by $\gamma$ or IC decay. $\gamma$s or IC electrons are emitted according to an exponential time distribution with a decay constant related to the half-life of the isomeric level. The following de-excitation to the ground state can occur via subsequent $\gamma$ or IC transitions.
\begin{table}[t]
	\centering
	\footnotesize
	\begin{tabular}{llc}
		\toprule
		$\beta^-$ & IC & $\gamma$/X-ray \\
		(keV) & (keV) & (keV) \\
		\midrule
		\multirow{3}{*}{$\beta^-$ (330)} & $e^-$ (278) & \multirow{3}{*}{$\gamma$ (106) , $\gamma$ (106) + X-ray} \\
		& $e^-$ (228) & \\
		& $e^-$ (210) & \\
		\midrule
		\multirow{3}{*}{$\beta^-$ (330)} & $e^-$ (8) & $\gamma$ (278) \\
		& $e^-$ (57) & $\gamma$ (228) \\
		& $e^-$ (75) & $\gamma$ (210) \\
		\midrule
		\multirow{4}{*}{$\beta^-$ (330)} & \multirow{4}{*}{$e^-$ (106)} & X-ray (99, 104, 116, 120) \\
		& & $\gamma$ (278) \\
		& & $\gamma$ (228) \\
		& & $\gamma$ (210) \\
		\bottomrule
	\end{tabular}
	\caption{Example of the main observed signatures. The first column is the $\beta^-$ transition to the metastable level. The next two columns on the right are the delayed transitions detectable by GeSparK detector. These transitions are the main de-excitation channels of the metastable level. Other transitions can also occur with lower probability, contributing to the total signal.}
	\label{tab:Transitions}
\end{table}
The LS detector allows to detect with high efficiency and good time resolution both the $\beta^-$ and IC/$\gamma$ electrons, while the HPGe detector is useful to detect the $\gamma$ or X photons frequently emitted as a consequence of the IC transitions.
The $\beta$-$\gamma$ coincidence capability of the GeSparK detector was exploited to select different decay channels in order to evaluate possible systematic uncertainties and to perform a reduction of the possible random coincidences. In Table \ref{tab:Transitions} the main observed signatures are reported. 
Measuring the delay between the two signals generated in the liquid scintillator from $\beta^-$ and IC/$\gamma$ electrons it is possible to construct the life distribution of the metastable levels that are populated by the observed beta decays. An exponential least squares fit on the obtained time difference distribution allows to achieve an accurate evaluation of the half-life.

\section{Half-life measurement}
\label{S:4}

\subsection{Source preparation}

A dedicated experiment was arranged in order to estimate the half-life of the metastable level.
To perform this measurement, a source of $^{239}\textrm{Np}$ was produced by neutron activation at the research reactor TRIGA Mark II at Applied Nuclear Energy Laboratory (LENA) of the University of Pavia (Italy), irradiating a sample of $^{238}\textrm{U}$ certified standard solution.
The total irradiated mass of $^{238}\textrm{U}$ was about 0.5~$\mu$g diluted in 2.5 mL of water.
Equation \ref{eq:Reaction} shows the neutron activation reaction:
\begin{equation}
^{238}\textrm{U} + n \rightarrow \:^{239}\textrm{U} \xrightarrow[]{\text{23.45 m}} \:^{239}\textrm{Np} + e^- + \overline{\nu}_e
\label{eq:Reaction}
\end{equation}

After six hours of irradiation in the Lazy Susan channel ($\phi_n\sim2\cdot10^{12}\frac{\textrm{n}}{\textrm{s}\cdot \textrm{cm}^2}$\cite{Chiesa1,Chiesa2}) the sample was dissolved in the liquid scintillator of GeSparK detector (Ultima Gold AB - Perkin Elmer) and sealed in the Teflon container in order to be measured with the $\beta$-$\gamma$ detector.

\subsection{Experimental measurement and data acquisition}

\begin{figure}[t]
	\begin{center}
		\includegraphics[width=0.48\textwidth]{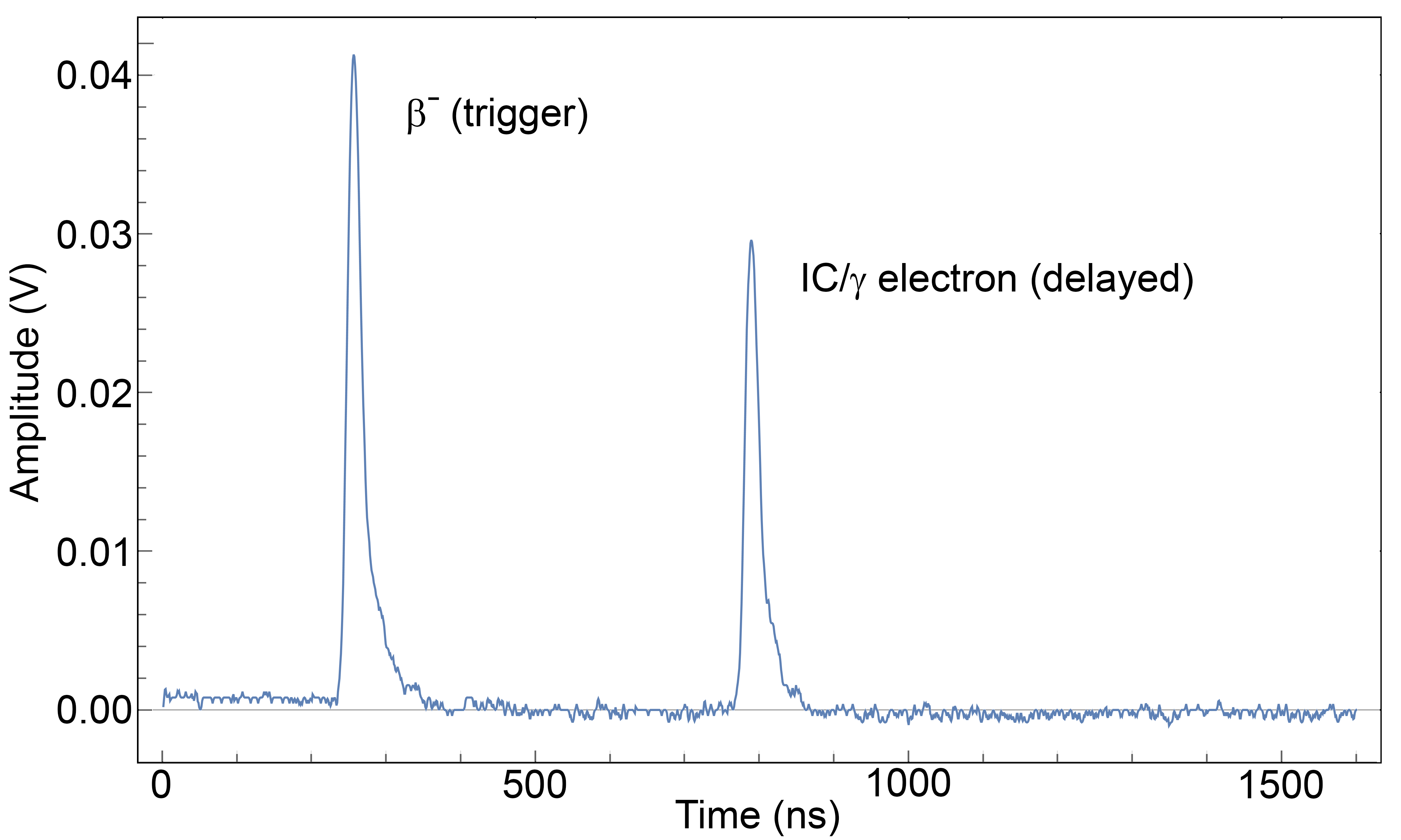}
		\caption{Example of the acquired signals from the LS detector. The first pulse, at around 250 ns, is the trigger one that is identified as the signal produced by the electron emitted in the $\beta^-$ decay to the metastable level. The second pulse is associated to the IC/$\gamma$ electron emitted in the delayed de-excitation cascade of that level.}
		\label{fig:signal}
	\end{center}
\end{figure}
\begin{figure}[t]
	\begin{center}
		\includegraphics[width=0.48\textwidth]{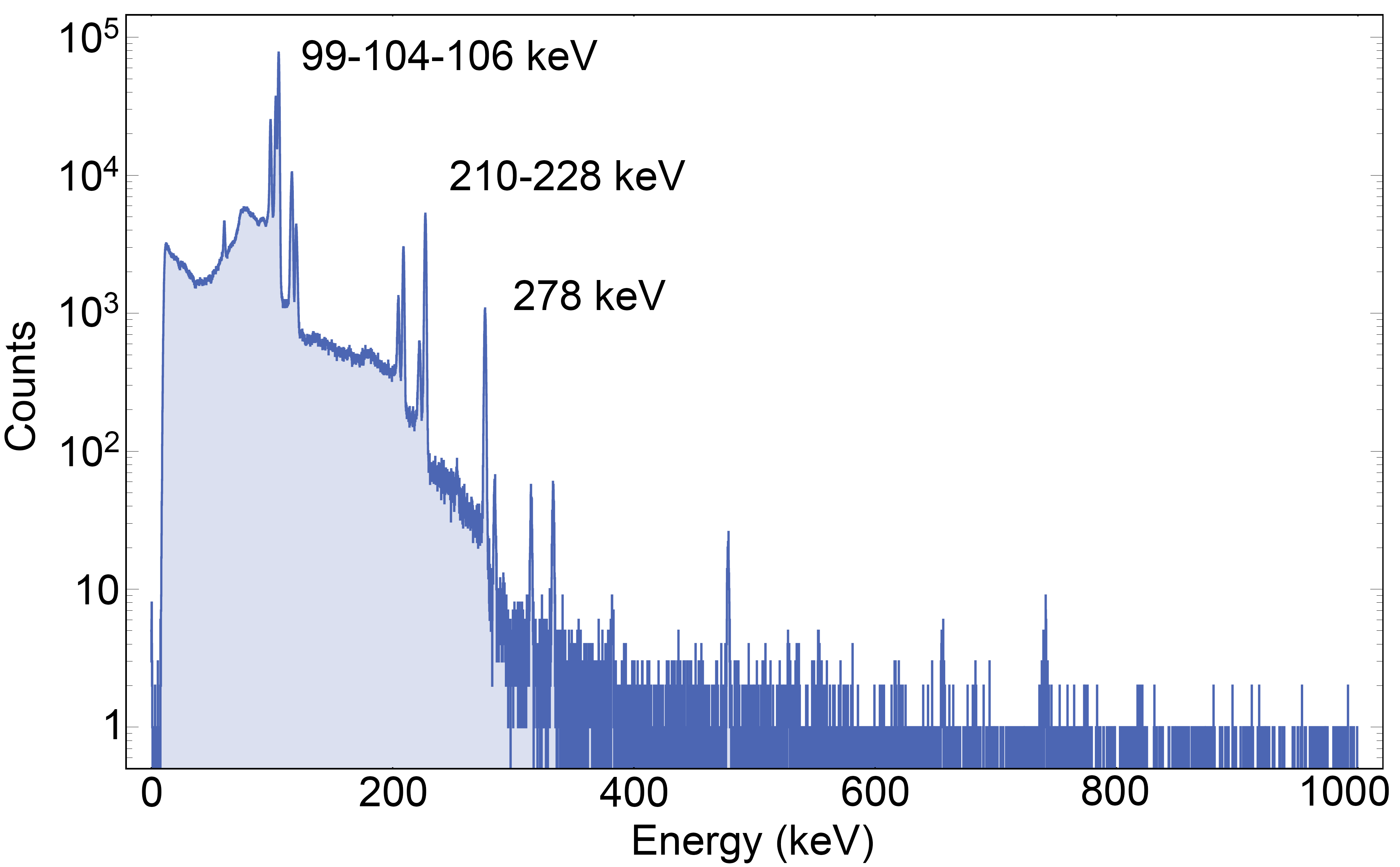}
		\caption{HPGe energy spectrum of gammas in coincidence with delayed LS events. The main $\gamma$ lines of $^{239}\textrm{Pu}$ are labeled in the plot.}
		\label{fig:HPGe}
	\end{center}
\end{figure}

In the GeSparK detector, when a charged particle releases its energy in the LS detector, an electronic pulse is sent from the PMT to the digital acquisition system (DAQ). The DAQ was set to digitize each triggered event from the PMT with a time division of 1\,ns and a time window ($\Delta t_{w}$) 1600\,ns wide, of which 270 ns are pre trigger. The width of the time window was set to more than seven half-lives of the $^{239}\textrm{Pu\,(391.6\,keV)}$ level in order to acquire the trigger event and a possible second event within this time interval. Figure \ref{fig:signal} shows an example of the LS acquired signals. The first pulse (trigger) is identified as the $\beta^-$ electron signal and the second one as the IC/$\gamma$ electron signal (delayed). In coincidence with the PMT pulse, also the HPGe signal is digitized in order to verify the presence of the coincident $\gamma$/X-ray emission. Figure \ref{fig:HPGe} shows the spectrum of the HPGe signals in coincidence with the LS pulses. Therefore, for each detected coincidence event the LS and HPGe detectors acquired data are stored.
The measure of the activated sample lasted 284 hours with a coincidence rate, at the measurement start, of about 150 Hz and a $^{239}\textrm{Np}$ source activity of 1050 Bq. 

\subsection{Analysis and results}

\begin{figure}[ht!]
	\begin{center}
		\includegraphics[width=0.48\textwidth]{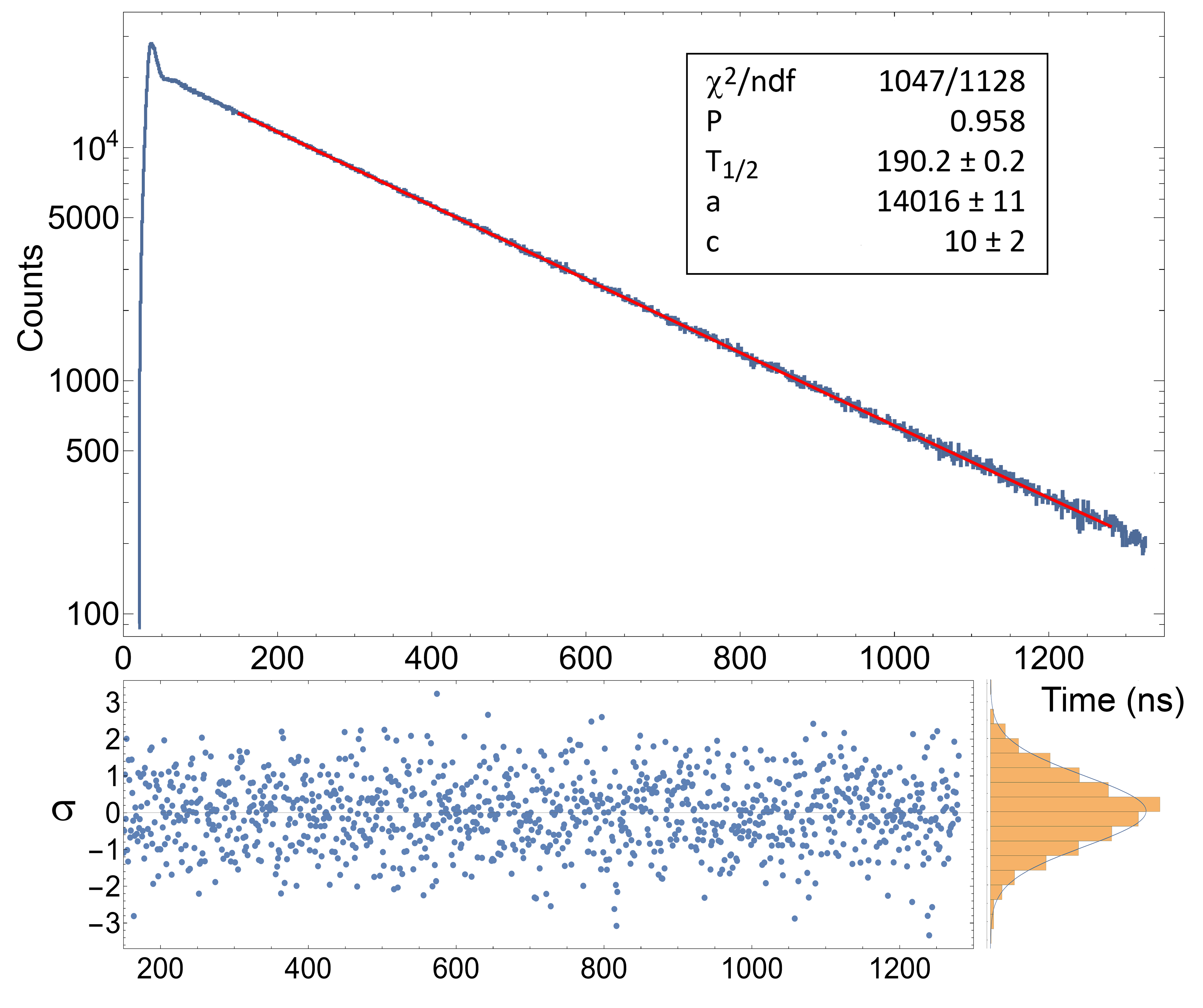}
		\caption{Top panel: distribution of the time differences, between $\beta^-$(trigger event) and IC/$\gamma$ electron (delayed event). Red line shows the best fit in the range 150\,ns - 1280\,ns, with a bin width of 1\,ns. Bottom panel: pull distribution.}
		\label{fig:fit}
	\end{center}
\end{figure}

An algorithm to perform the automatic detection of the pulses and the calculation of the relative time distance in each LS acquired window was developed. Figure \ref{fig:fit} shows the resulting distribution of the time differences between the $\beta^-$ trigger events and the delayed IC/$\gamma$ electrons. 

The fit of the time distribution was performed with a function defined by a decreasing exponential plus a constant. The analytical form of the fit function is the following:
\begin{equation}
	f(t)=a\cdot e^{-\frac{ln{2}\cdot t}{T_{1/2}}}+c
	\label{eq:fit}
\end{equation}
where \textit{a} is the amplitude of the exponential term and \textit{c} is the flat component to account for random coincidences in the approximation $R\cdot\Delta t_{w}<<1$ (R is the source rate). To reduce the contribution of random events generated from interference nuclei, activated during the irradiation, only LS events in coincidence with a $\gamma$ ray below 300\,keV were considered in the analysis. This was possible because beyond that energy value the contribution of $^{239}\textrm{Np}$ signals events is negligible with respect to the background.

The distortion at the beginning of the distribution of Figure \ref{fig:fit} is due to the pile up of the trigger event with the delayed one. This affects both the determination of the delays and the evaluation of the pulse amplitudes.
In order to exclude the events that are affected by pileup, the lower limit of the fit was set at 150\,ns, according to the timing features of the LS pulses (pulse width $\sim$ 100~ns). The upper limit of the adaptation has been set at 1280\,ns in order to remove the signals acquired at the end of the time window, since it is not sure to correctly measure their properties.

The best fit of the distribution and the fitting parameters are shown in Figure \ref{fig:fit}. The goodness-of-fit is satisfactory and the pull distribution in the bottom panel shows a good agreement between data and model. The obtained $\chi^2/\textrm{ndf}$ (0.928) and the corresponding probability (0.958) show a very good agreement between the data distribution and the fit function.
The best estimation of the $^{239}\textrm{Pu}$ half-life is $190.2\pm0.2\,$ns.

\subsection {Analysis of the systematic uncertainties}
During the analysis process some possible sources of systematic errors for the determination of the $\textrm{T}_{1/2}$ of the 391.6~keV metastable level were identified and their contribution was evaluated. These are:

\begin{itemize}
    \item $T_{1/2}[^{239}\textrm{Pu\,(285.5\,keV)}]$=1.12 ns
    \item ADC clock accuracy
    \item Histogram binning 
    \item Fit threshold
\end{itemize}

The presence of the 285.5 keV level in the decay scheme of $^{239}\textrm{Pu}$ (Figure 1) could introduce a systematic since this level, energetically below the 391.6 keV, is also a metastable state with a known half-life of 1.12 ns. Some events detected during the measurement are characterized by decay cascades that involve both these levels. In this case the time delay from the trigger event ($\beta^-$) and IC/$\gamma$ electron is shifted by a quantity related to the $\textrm{T}_{1/2}$ of 285.5 keV) level. The resulting time delay distribution of this specific events is described by the convolution of two exponential functions, whose decay constants are given by the mean life of the two levels, as reported in the following equation:
\begin{equation}
	\left(Exp(\tau_L)*Exp(\tau_S)\right)(t) = \frac{\tau _L \tau _S \left(e^{-\frac{t}{\tau _L}}-e^{-\frac{t}{\tau _S}}\right)}{\tau _L-\tau _S}
	\label{eq:convDistribution}
\end{equation}
where $\tau_L$ ($190/\ln(2)$ ns) and $\tau_S$ ($1.12/\ln(2)$ ns) are the mean lives of the 391.6 keV and 285.5 keV metastable levels respectively.
Since $\tau_L\!>>\!\tau_S$, for $t>>\tau_S$ the contribution of the fastest exponential term is negligible. This assumption is verified on the analysis since we set the lower limit of the fit interval at 150 ns, that is much higher than 1.12 ns.
Another proof was obtained by a toy Monte Carlo simulation. In this case the delays produced by the two metastable levels were simulated by generating a random number according to their exponential distribution. A fit of the resulting distribution was performed excluding the first 150 ns, obtaining a result perfectly compatible with the longer mean life.

A further source of systematic error could be the accuracy of the ADC clock. In accordance with the warranted specifications of the ADC (National Instrument mod. PXI-5153), this contribution was evaluated in tens of picoseconds, thus negligible.

\begin{figure}[t]
	\begin{center}
		\includegraphics[width=0.48\textwidth]{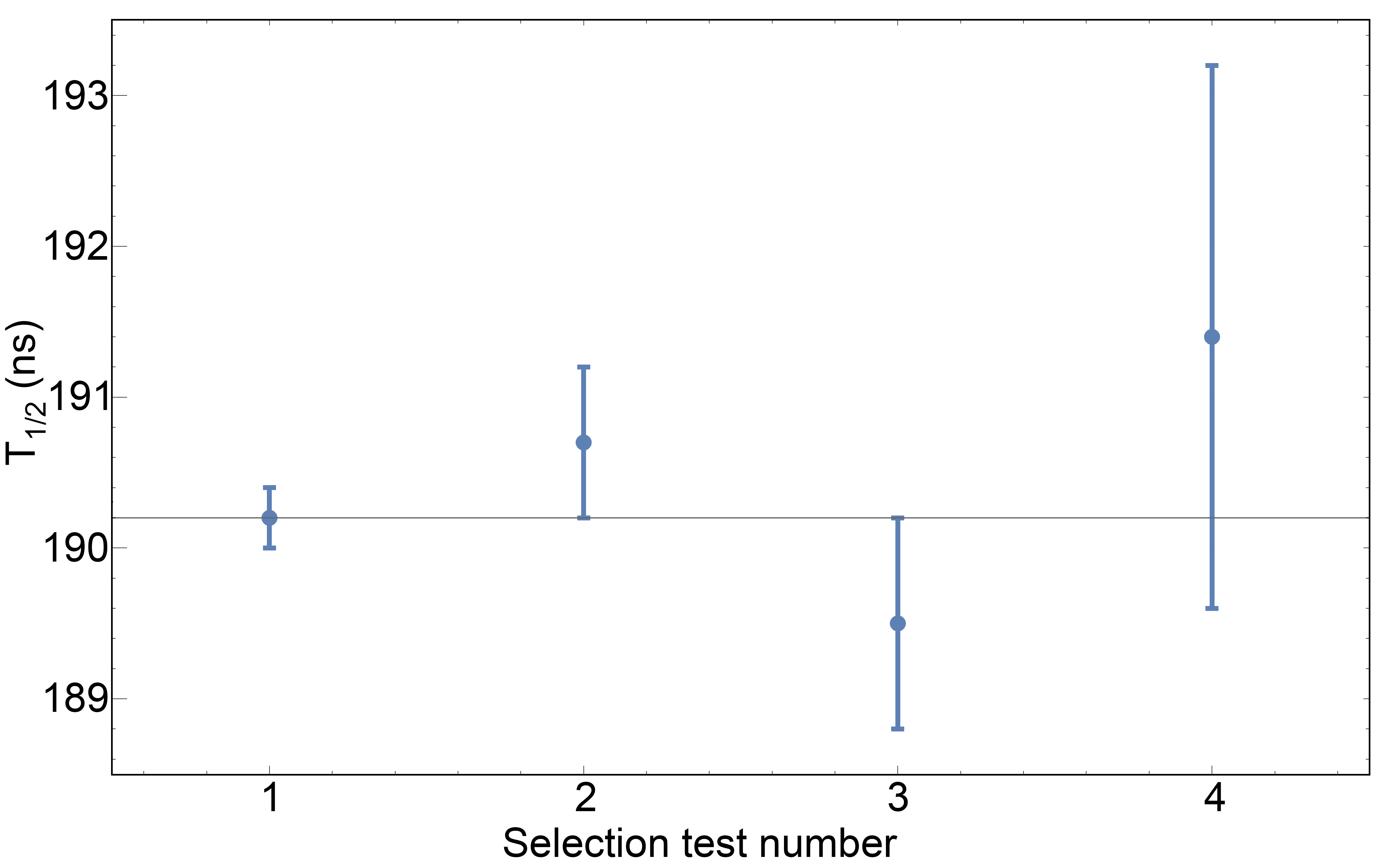}
		\caption{Mean life obtained by fitting different time distribution constructed selecting a particular decay channel using the HPGe coincidence. The test 1 is obtained selecting all the gammas below 300 keV. The test 2, 3 and 4 are performed selecting respectively the 106 keV, 104 keV and 228 keV peaks.}
		\label{fig:selectiontest}
	\end{center}
\end{figure}

Finally, the distribution of the time differences in Figure 4 was fitted for different choices of the binning in the histogram and fitting threshold between 150\,ns and 650\,ns. In both cases the variations of the fit result are negligible with respect to the statistical error associated to the measurement.
Thanks to these considerations it can be stated that statistics dominate the uncertainty of the final result. Moreover, this is also a test to study the presence of other radioactive contaminants that would produce different half-life estimations by changing the fit threshold. 

In order to bring out other systematic errors not considered in the above list, a validation of the obtained result was performed. The presence of the HPGe detector in the experimental setup allows to select with a good energy resolution gamma or X photons. By forcing an energy selection for $\gamma/\textrm{X-ray}$ in the analysis of the acquired events, it is possible to identify specifically observed decay sequences. Figure \ref{fig:selectiontest} shows the half lives obtained from the different selections. The first point is the result achieved in the previously reported analysis selecting all the gammas with energy below 300 keV. The points 2, 3 and 4 were instead obtained by selecting respectively 106 keV, 104 keV and 228 keV energy emissions.
The selections are representative of different types of transitions in the decay scheme (Figure \ref{fig:decayscheme}). Since the results obtained from the fits are compatible within one standard deviation, it is possible to conclude that the effect of selecting a specific decay sequence is negligible (e.g. presence of 285.5 keV metastable level).
This test also demonstrates the possibility of using all the gammas below 300 keV in order to increase the statistics of the measurement.

\section{Conclusions}
\label{S:5}

In this work the measurement of the half-life of the metastable level at 391.6~keV of the $^{239}\textrm{Pu}$ is presented. 
The novel measurement technique, which exploits the delayed coincidences generated between $\beta^-$ decay and IC/$\gamma$ electron emissions, allowed to measure the half-life of the isomeric nuclear states. The applied measurement technique proved to be a good tool to perform similar measurements for nuclei that have a similar decay sequence and that are of particular interest in the field of nuclear physics.

The dedicated analysis performed in this work, allowed to achieve the best results for $T_{1/2}$ of the 391.6~keV level: $190.2 \pm 0.2\,\textrm{ns}$. This value is statistically compatible with the best-known value \cite{Engelkemeir} but with a factor 20 smaller uncertainty and it represents an advancement in the knowledge of the $^{239}\textrm{Pu}$ nuclear levels.

The decay of $^{239}\textrm{Np}$ on $^{239}\textrm{Pu}$ has an important application in neutron activation analysis for $^{238}\textrm{U}$ quantification, crucial in material selection for rare events physics experiments. In fact, using a $\beta/\gamma$ coincidence detector allows to reduce the background, but the sensitivity could still be limited by interferent $\beta$ decaying isotopes activated in the samples. In that case, the time analysis of delayed events produced by the 391.6 keV metastable level is crucial to disentangle the $^{239}\textrm{Np}$ signals, thus increasing the sensitivity in the search for $^{238}\textrm{U}$ contaminations. A similar approach has already shown that it could be a very effective way to increase sensitivity below $10^{-13}$ g/g \cite{Goldbrunner}.


\bibliographystyle{unsrt2authabbrvpp}
\bibliography{Bibliography}

\end{document}